\documentstyle[11pt,moriond,psfig]{article}

\def\Journal#1#2#3#4{{#1} {\bf #2}, #3 (#4)}
\def\Nat{{\em Nature}}
\def\AJ{{\em Astron. J.}}
\def\APJ{{\em Astrophys. J.}}
\def\APJL{{\em Astrophys. J. Lett.}}
\def\MNRAS{{\em Mon. Not. R. Astron. Soc.}}
\def\ARAA{{\em Ann. Rev. Astron. Astrophys.}}

\def\fun#1#2{\lower3.6pt\vbox{\baselineskip0pt\lineskip.9pt
  \ialign{$\mathsurround=0pt#1\hfil##\hfil$\crcr#2\crcr\sim\crcr}}}

\def\etal{{\rm et al.}}

\begin{document}

\noindent To appear in: {\sl Fundamental Parameters in Cosmology},

\noindent XXXIII Recontres de Moriond, eds. Y. Giraud-H\'eraud et al., 

\noindent Gif sur Yvette: Editions Fronti\`eres, in press (1998)

\vspace*{3cm}
\title{HIGH-REDSHIFT QUASARS AS PROBES OF PRIMORDIAL LARGE-SCALE
STRUCTURE AND GALAXY FORMATION}

\author{S. G. DJORGOVSKI}

\address{Palomar Observatory, Caltech, Pasadena, CA 91125, USA}

\maketitle\abstracts{
Several arguments suggest that quasars at $z > 4$ may be in the cores of future
giant ellipticals, and forming at the very highest peaks of the primordial
density field.  A strong bias-driven primordial clustering is then expected in
these fields, which are naturally interpreted as the cores of future rich
clusters.  High-redshift quasars can thus be used as markers of some of the 
earliest galaxy formation sites.  Recent discoveries of (proto)galaxy 
companions of $z > 4$ quasars, and hints of strong clustering among high-$z$
quasars give some support to this idea.  The frequency of companions is much
higher than is expected for a general field, or the upper limits for comparable
redshifts in the HDF.  Clustering of quasars themselves at $z > 3$ or 4 may
reflect the primordial large-scale structure.
}

\section{Introduction}

Primeval galaxies (PGs) have now been found, in a variety of forms and
environments, using a variety of techniques: 
narrow-band Ly$\alpha$ imaging~\cite{Cow97},
Lyman-break method~\cite{Ste96}, 
as DLA absorbers~\cite{Djo96}, 
as quasar companions~\cite{Djo85,Hu96},
as gravitationally lensed objects~\cite{Fra97},
etc.  Statistical studies of the HDF~\cite{Madau}
have outlined a global history of galaxy formation in the general field. 
Moreover, we are starting to see evidence ~\cite{Ste98}
for large-scale structure formation at high redshifts, i.e., $z > 3$. 

However, any single discovery method may bias the sample of objects found, and
formative histories of galaxies in different environments may vary
substantially.  For example, galaxies in rich clusters are likely to start
forming earlier than in the general field, and studies of galaxy formation in
the field may have missed possible rare active spots associated with rich
protoclusters.  Diverse and complementary paths towards discoveries of first
galaxies are still very much worth pursuing.  Our group is pursuing the use 
of quasars at $z > 4$ as markers of the early galaxy formation sites, by
looking for PGs associated and/or clustered with quasar host galaxies
themselves ($z_{gal} \approx z_{QSO}$). 

It is now believed that galaxy formation is an extended series of processes,
which may have a broad peak at $z \sim 2 - 3$.  However, it is of a
considerable importance to catch the $first$ galaxy-size structures forming at
high redshifts, i.e., the very onset of massive galaxy formation.

Any galaxies
at $z > 4$ must be still very young, simply on account of the timing: they can
be at most a few percent of the present galaxian age, or $\sim 0.5 - 1$ Gyr,
for any reasonable cosmology.  The evolutionary status of galaxies at $z \sim
2$ or 3 is more ambiguous, with possible physical ages up to $\sim 2 - 3$ Gyr,
or $\sim 10 - 20$\% of their present age.  Going to $z > 4$ removes this age
ambiguity.

We are conducting an observational program to discover and study such
protogalaxies in the earliest stages of formation, clustered with the known
quasars at $z > 4$.  This pushes beyond the territory now explored in the HDF
and elsewhere, at the epochs when the universe was only a few percent of its
present age. 

\section{Quasars as Cores of Massive Protogalaxies}

There is currently very little known about normal galaxies and their
progenitors at $z > 4$.  While the current theoretical belief \cite{MirEsc}
is that some subgalactic structures ($M \sim 10^6 - 10^8 M_\odot$) may start
forming at $z \sim 6 - 10$, more massive ($M \sim 10^{11} - 10^{12} M_\odot$)
progenitors of normal galaxies today are unlikely to be in place before $z \sim
4 - 5$. 

Some of these first massive protogalaxies may be the hosts of quasars at
$z > 4$.  While AGNs by themselves are problematic as a probe of galaxy
formation and evolution, they may still provide useful pointers to the sites
of early galaxy formation.  The relation between high-$z$ AGN and galaxy
formation remains unclear.  The comoving density of quasars seems to track very
well the inferred history of star formation in galaxies \cite{Madau},
and both follow the merger rate evolution predicted by hierarchical
structure formation scenarios.  The same kind of processes, dissipative merging
and infall, may be an important trigger of both star formation and the AGN
activity.  

Indeed, most or all ellipticals and massive bulges at $z \sim 0$
seem to contain central massive dark objects suggestive of an earlier quasar
phase \cite{KormRich}.  
Masses of these
putative ``dead quasars'' also correlate with the luminous old stellar mass of
the host galaxy, again suggestive that they formed at about the same time
and through the same sequence of early dissipative merging and starbursts.

The very existence of numerous quasars at $z > 4$ poses a timing problem, and
it is likely that they are situated in the massive parent structures (future
host galaxies) which are among the first objects to form \cite{Tur91}.
Furthermore, high metallicities (up to $10 \times ~Z_\odot$!) observed in $z >
4$ quasars \cite{HamFer93} 
are indicative of a considerable chemical
evolution involving several generations of massive stars in a system massive
enough to retain and recycle their nucleosynthesis products, e.g., comparable
to the cores of giant ellipticals.  Metallicities and abundance patterns in the
intracluster x-ray gas at lower redshifts are also suggestive of an early,
rapid star formation and enrichment phase in the rich cluster progenitors at
high redshifts \cite{LoewMush96}.

It is perfectly possible that $every$ young elliptical or a massive bulge
undergoes an early quasar phase as it is forming the bulk of its stars.  While 
quasar activity can happen at any redshift if there is a central engine in
place and a supply of fuel available, the situation at high redshifts is
special: these objects must be young on account of the short time scales.
Quasars at $z > 4$ can thus be naturally interpreted as being in the cores 
of future massive ellipticals.

In general, the most massive density peaks in the early universe are likely to
be strongly clustered \cite{Kai84} and thus the first galaxies may be forming
in the cores of future rich clusters.  It then makes sense to look for other
galaxies with or without AGN, forming in the immediate vicinity of known $z >
4$ quasars. 

\section{Quasar Clustering at z $>$ 3}

There is a possibility that some quasar-marked protoclusters at high redshifts
have already been discovered.  Schneider, Schmidt \& Gunn (SSG) \cite{SSG94}
present the results of a grism search for quasars at $2.7 < z < 4.9$,
covering 61.47 deg$^2$.  In this volume, they define a $complete$ sample of 90
quasars, which they used to measure the evolution of the quasar luminosity
function.  However, they have found something remarkable: among their complete
sample of 90 quasars, there are {\it 3 close pairs}, all at $z > 3$:
PC 0951+4637 A+B:
$\langle z \rangle = 3.223$, $\Delta z = 0.005$, $\Delta \theta = 410$";
PC 1314+4748 A+B:
$\langle z \rangle = 3.355$, $\Delta z = 0.004$, $\Delta \theta = 364$"; and
PC 1643+4631 A+B:
$\langle z \rangle = 3.810$, $\Delta z = 0.041$, $\Delta \theta = 198$".

The corresponding comoving separations are $\sim 3 - 5$ Mpc, i.e., comparable
to the sizes of galaxy clusters.  The volume covered by the SSG survey is huge,
and for each quasar pair, the probability of a random occurrence in this volume
is $\sim$ a few $\times 10^{-5}$.  The pairs are detected independently. 
Thus, the joint probability of finding 3 pairs this close in this survey is $P
\sim 10^{-13}$!  This clearly cannot be due to a chance.  Assuming a standard 
power-law clustering with slope $\gamma = -1.8$, the implied comoving
clustering length is of the order of $50 h^{-1}$ Mpc, i.e., comparable to that
of the richest Abell clusters today.  It thus appears very likely that these
quasar pairs mark sites of rich protoclusters of galaxies.  

While sporadic evidence for clustering of quasars at high redshifts has been
presented in the past, this was the first time that a complete and well-defined
sample of quasars was available for this purpose, enabling meaningful
statistical arguments.  Clustering of absorbers and their companion galaxies
has also been detected in several instances, and of Lyman-break galaxies as
well: we know that some large scale structure is in place at $z \sim 2 - 3$.

Studies of the clustering of quasars for the entire SSG sample have been
done by two groups \cite{Step97,Kund97}.
Both have found some evidence for clustering, which is dominated by these
pairs.  Previous studies of quasar clustering (typically out to $z \sim 2$ or
so) have generally found that their clustering amplitude decreases with
redshift, presumably reflecting the growth of the large-scale 
structure \cite{HartSchad}.
Here we seem to see the reversal of this trend: a sudden (and perhaps rather
dramatic) increase in the clustering strength of quasars at $z > 3$.  Similar
effect was also found by La Franca et al. \cite{LaFra98}, 
in an analysis of a completely independent sample of objects at somewhat lower
redshifts. 

This can be interpreted as an evidence for biasing: these quasar pairs may be
marking some of the highest peaks of the density field at that epoch, 
probably marking the regions which will evolve into rich galaxy clusters today.

We searched for other objects at the same redshifts in these quasar-pair fields
using narrow-band imaging centered on the redshifted Ly$\alpha$ line to select
star-forming galaxies and faint AGN at the same redshifts.  Our preliminary
spectroscopic follow-up confirmed several objects, including both Ly$\alpha$
emission galaxies, and previously unknown, faint quasars \cite{Djo98}.  
These findings are certainly consistent with the idea that these quasar pairs
may be marking sites of future rich clusters \cite{Djo93}. 

It then makes sense to push to even higher redshifts, and look for objects
clustered with known quasars at $z > 4$.

\section{New Discoveries of Quasar Companions at z $>$ 4}

This approach has been proven to work.  The first galaxy discovered at $z > 3$
was a quasar companion (PKS 1614+051 at $z =3.215$) \cite{Djo85}, and many
other examples have been found since then at $z \sim 2 - 3$.
This method is now starting to yield non-AGN galaxies at $z > 4$.

A galaxy companion of BR 1202--0725 at $z = 4.695$ has been discovered
independently and at the same time by two groups \cite{Djo95,Hu96},
and confirmed spectroscopically \cite{Petit96}.
No high-ionization lines detected in its spectrum, but the close
proximity of this object to the QSO suggests that its strong Ly$\alpha$
emission may be powered by the QSO radiation field, rather than by star
formation.  Furthermore, a dusty companion object has been found in the same
field \cite{Omont95,Ohta95}.  
Two companion galaxies were also found \cite{HuMcM}
in the field of the quasar BR 2237--0607 at $z = 4.55$.

\begin{figure}
\centerline{\psfig{file=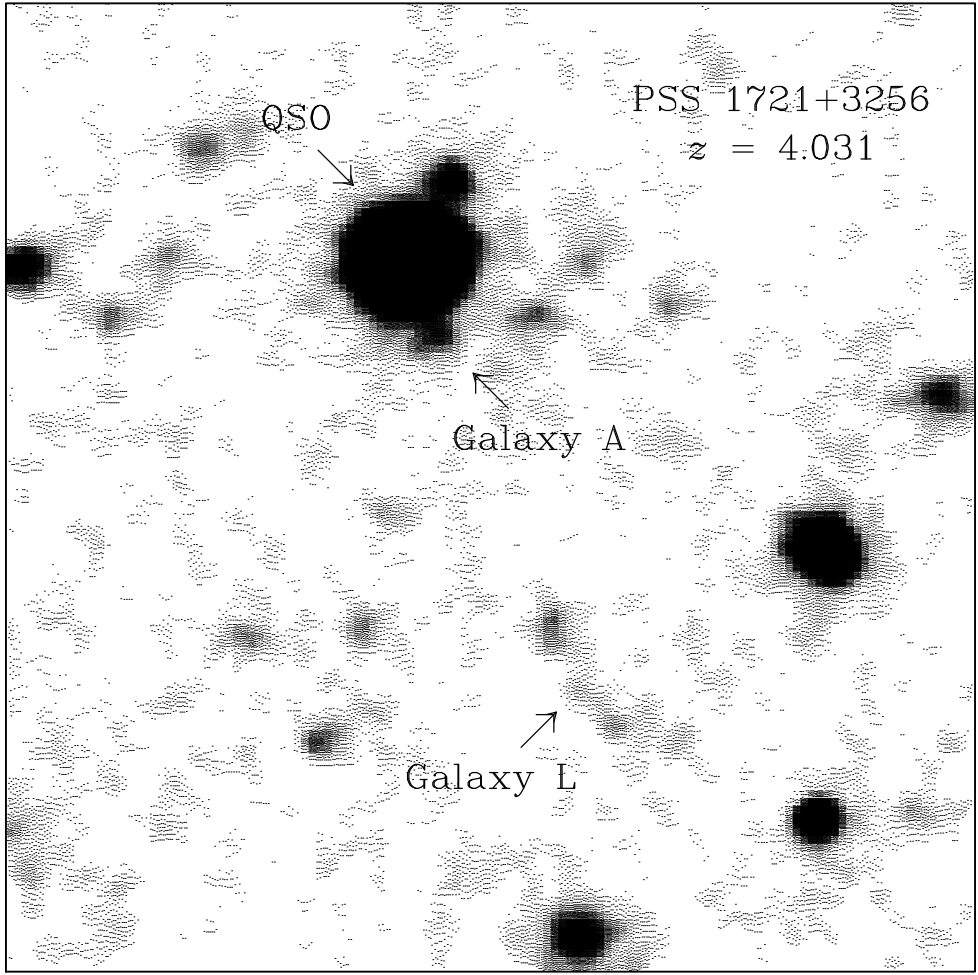,width=5.5in}}
\caption{
Deep $R$ band Keck image of the field of PSS 1721+3256, a quasar at $z =
4.031$.  The field shown is 27.5 arcsec square.  Galaxy L is an $R \approx
25^m$ object at the QSO redshift; galaxy A is probably also at the same
redshift.
}
\label{fig:PSS1721}
\end{figure}

\begin{figure}
\centerline{\psfig{file=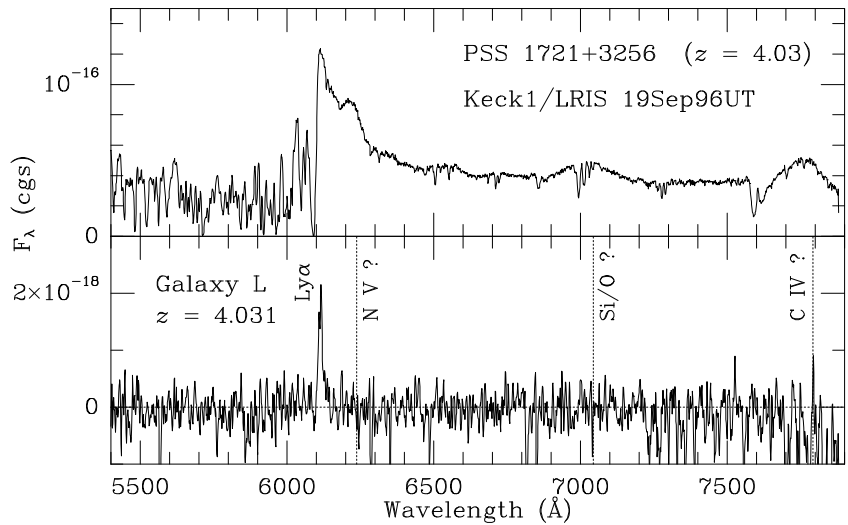,width=4.0in}}
\caption{
Keck spectra of the quasar PSS 1721+3256 and its companion galaxy L at $z =
4.031$.  Note the complete absence of high-ionization lines in the galaxy's
spectrum, suggesting that it is powered by star formation, rather than by
an AGN.}
\label{fig:spec1721}
\end{figure}

There are now several other cases.  For example,
we have recently found \cite{Djo98}
an apparently normal galaxy companion of the quasar PSS
1721+3256 at $z = 4.031$ 
(Fig.~\ref{fig:PSS1721}).
It is an $R \approx 25^m$ galaxy seen 13 arcsec in projection from the QSO
(corresponding to $\sim 450$ comoving kpc, or $\sim 90$ proper kpc at that
redshift, for a reasonable range of cosmologies), sufficiently far as to be
relatively unaffected by the QSO radiation field.  Its inferred continuum
luminosity is $\sim L_*$, and its spectrum 
(Fig.~\ref{fig:spec1721})
shows no trace of an AGN:
no high ionization lines, relatively weak Ly$\alpha$ (an order of magnitude
less than the companion of BR1202--0725) with $W_{\lambda ~rest} \approx 25$
\AA.  The inferred star formation rate, both from the Ly$\alpha$ emission line
and the UV continuum at $\lambda_{rest} = 1500$\AA, is SFR $\approx 4$ to $5
~M_\odot$/yr. Its properties are then exactly what one may expect from a PG
powered by a mild star formation.  Another possible Ly$\alpha$ companion galaxy
in the same field (galaxy A) awaits a further spectroscopic confirmation.

\begin{figure}
\centerline{\psfig{file=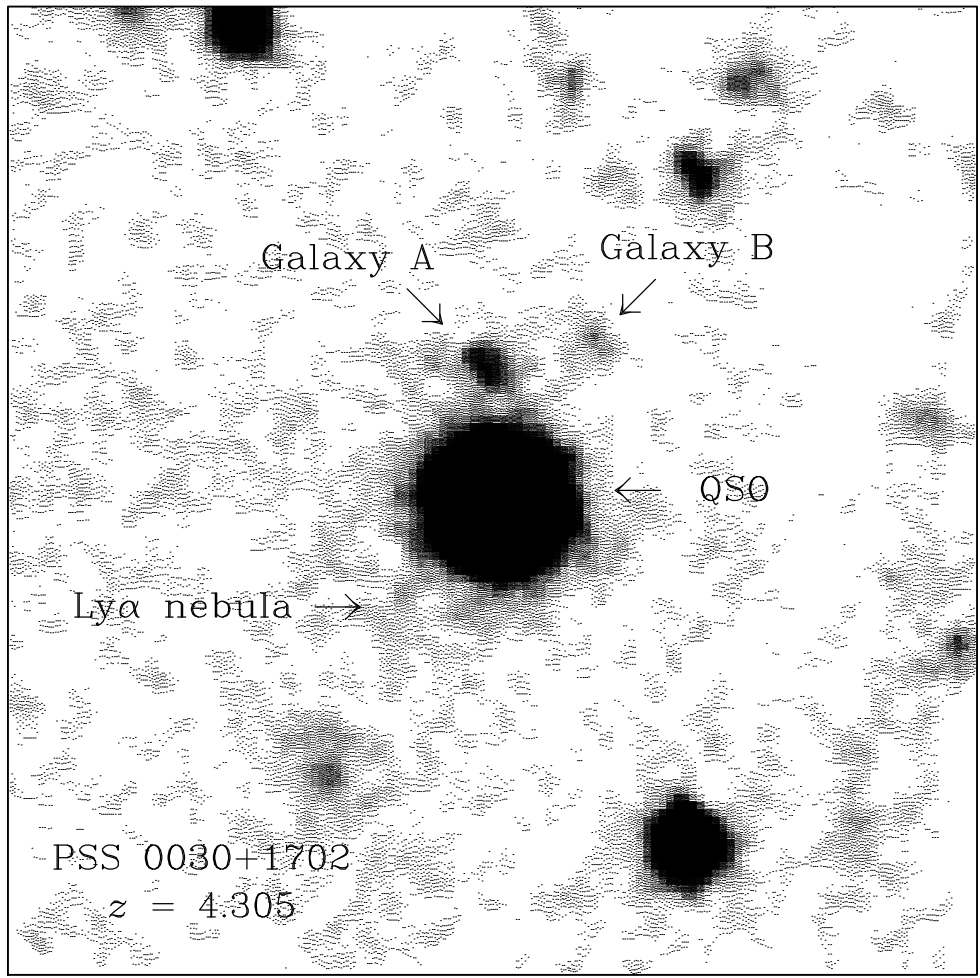,width=5.5in}}
\caption{
Deep $R$ band Keck image of the field of the quasar PSS 0030+1702,
at $z = 4.305$.  The field shown is 27.5 arcsec square.
Galaxy A is an $R \approx 24^m$ object at the QSO redshift; galaxy B ($R
\approx 25^m$) may also be at the same redshift.  There is a pure Ly$\alpha$
line emission nebula on the opposite side of the quasar, probably a part of the
still gaseous QSO host galaxy ionized by its UV emission.
}
\label{fig:PSS0030}
\end{figure}

Another very interesting case is PSS 0030+1702 at $z = 4.305$ 
(Fig.~\ref{fig:PSS0030}).
The QSO has at least one, and
possibly two Ly$\alpha$ companion galaxies within 10 arcsec from the line of
sight \cite{Djo98}.  
Their observed parameters are similar to those of the galaxy companion
of PSS 1721+3256, described above: no high-ionization lines in their spectra,
relatively weak Ly$\alpha$, continuum luminosity $\sim 1 - 1.5 ~L_*$, SFR $\sim
10 \pm 5 ~M_\odot$/yr (both from the Ly$\alpha$ line, and the UV continuum
flux).  We also conclude that these objects are most likely powered by star
formation. 

We have also found \cite{Djo98}
similar cases of protogalaxy companions of quasars
PSS 0117+1552 at $z = 4.275$,
PSS 0248+1802 at $z = 4.465$,
PSS 1048+4407 at $z = 4.45$,
PSS 1057+4555 at $z = 4.12$,
PSS 1253--0228 at $z = 4.01$,
and GB 1713+2148 at $z = 4.01$;
several other cases which require more observations at this point.
Overall, the intrinsic properties of these quasar companion galaxies (their
luminosities, SFR, etc.) are very similar to those of the Lyman-break selected
populaton at $z \sim 3$, except of course for their special environments and
higher look-back times.

In addition to these galaxies where we actually detect (presumably starlight)
continuum, we sometimes see pure Ly$\alpha$ emission line nebul\ae\ within 
a few arcsec from the quasars, with no detectable continuum at all. One example
is PSS 0030+1702.  The  Ly$\alpha$ flux from the nebula exactly what may be
expected from photoionization by the QSO, with $L_{Ly \alpha} \approx 2 \times
10^{43}$ erg/s.  These nebul\ae\ may be parts of still gaseous protogalaxy
hosts of the quasars themselves.  We can thus see and $distinguish$ both the
objects powered by the neighboring QSO, and objects which by all signs appear
to be ``normal'' PGs in their vicinity. 

\section{Evidence for Protoclusters in Highly Biased Regions?}

The median projected separations of these objects from the quasars are $\sim
100 h^{-1}$ comoving kpc, an order of magnitude less than the comoving r.m.s.
separation of $L_*$ galaxies today, but comparable to that in the rich cluster
cores.  The frequency of QSO companion galaxies at $z > 4$ also appears to be
an order of magnitude higher than in the comparable QSO samples at $z \sim 2
- 3$, the peak merging epoch; galaxy interactions alone are thus probably not
the whole story here.  The implied average star formation density rate in these
regions is some 2 or 3 orders of magnitude higher than expected from the limits
estimated for these redshifts \cite{Madau} for $field$ galaxies.

\begin{figure}
\centerline{\psfig{file=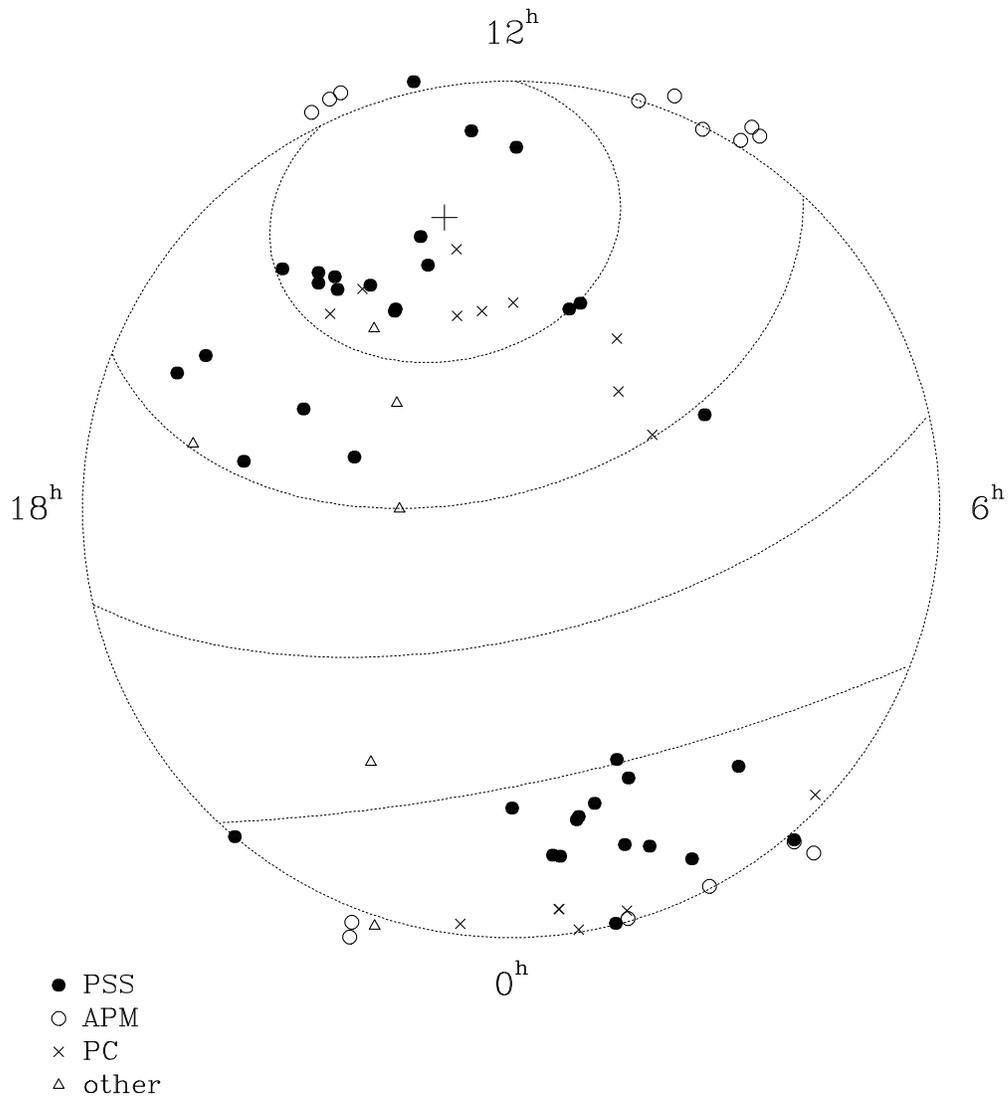,width=6.0in}}
\caption{
The distribution of known $z > 4$ quasars (as of May 1998) on the northern and
equatorial sky.  The majority are from the uniform, but still very incomplete
Palomar DPOSS survey (PSS) \protect\cite{Kennefick,Djo98} ; 
others are from the APM survey \protect\cite{APMQ} ,
from the Palomar transit surveys (PC) \protect\cite{SSG94} ,
and a few others selected in radio, x-rays, colors, or serendipity.
Several close pairs and triplets are seen, sometimes with components found
by independent surveys.  The dotted lines indicate the circles of equal
Galactic latitude, $b_{II} = -30^\circ, 0^\circ, +30^\circ$, and $+60^\circ$,
and the cross marks the North Galactic pole.
}
\label{fig:qsoclus}
\end{figure}

While this is still very preliminary, this ``excess'' may be an observable
manifestation of biasing, i.e., the expected clustering of the highest density
peaks.  A generic expectation in most models of galaxy formation, including the
standard CDM cosmogony, is that first objects forming at high redshifts should
be strongly clustered: they would presumably form in the highest peaks of the
density field, which are clustered {\it ab initio} \cite{Kai84,EfstRees88}.
The same explanation applies to the redshift ``spikes'' at $z \sim 3 - 3.5$ 
found by Steidel \etal ~\cite{Ste98}.
What we are finding may be even denser, and thus much rarer peaks at $z > 4$. 
Because they are rare, we use quasars as markers of sites where some structure
is already forming, in order to increase our chances.  Doing a ``pure deep
field'' approach at these redshifts would be much harder and less likely to
succed than at $z \sim 3$. 
These are obviously very special (highly biased) spots in the early universe.

Quasars at $z > 4$ may thus be interpreted as being in the metal-rich cores of
young, massive ellipticals, and possibly mark sites of future rich clusters of
galaxies.  The discoveries of their companion galaxies may support this idea:
these may be ``normal'' PGs in the cores of future rich clusters, in the
earliest stages of formation.  These are obviously very special spots in the
early universe, and they present a great opportunity to study galaxy and
cluster formation, in an enviroment deliberately different from the general
field (e.g., the HDF).  

On an even larger scale, there may be some evidence for excess power on
supercluster scales, as probed by quasars at $z > 4$ (Fig.~\ref{fig:qsoclus}).
While the numbers of objects are still relatively small, we are finding an
unexpectedly large frequency of QSO pairs and triplets (and even one quartet!)
with typical comoving separations of $\sim 100 h^{-1}$ Mpc \cite{Djo98}.
The statistical significance of this effect is still difficult to quantify,
due to the heterogeneity of some of the data, and the patchy sky coverage.
The effect may be spurious, e.g., due to the variable depth of the QSO
surveys or other selection effects, or it could be real.  If it is real,
it may be due to the actual clustering of quasars, or to some other physical
effect, e.g., a patchy gravitational magnification by the foreground
large-scale structure. 

If this is a real clustering signal, its implications would be profound.
It would be the first detection of a primordial large-scale structure, seen
by its highest, highly biased peaks containing quasars, only a few hundred
physical Mpc away from the CMBR photosphere.  As the data improve, both for
the QSO samples and the CMBR measurements, it would be intriguing to see if
any correlation can be found between these quasar-marked structures and
the CMBR fluctuations behind them.

Further work is needed in order to better establish and quantify these
findings.  If the evidence holds, it would represent a powerful confirmation of
our basic ideas about biased galaxy formation.  In any case, we are begining to
probe the earliest phases of galaxy and large-scale structure formation at high
redshifts.

\section*{Acknowledgments}
Various parts of this work are done in collaboration with 
S. Odewahn,
R. Gal,
R. de Carvalho,
L. Ferrarese,
M. Pahre,
K. Banas, 
and several others; their efforts and contributions are gratefully
acknowledged.  I also wish to thank the staff of Palomar and Keck observatories
for their expert help during many observing runs.
This work was supported in part by the Bressler Foundation and the Norris
Foundation.  I thank the organizers for a partial travel support.

\end{document}